\documentclass[12pt]{article}
\usepackage{epsfig}

 \newcounter{mytime}
\def\figcap{\section*{Figure Captions\markboth
        {FIGURECAPTIONS}{FIGURECAPTIONS}}\list
        {Figure \arabic{enumi}:\hfill}{\settowidth\labelwidth{Figure
999:}
        \leftmargin\labelwidth
        \advance\leftmargin\labelsep\usecounter{enumi}}}
 \relax
\newskip\humongous \humongous=0pt plus 1000pt minus 1000pt
\def\caja{\mathsurround=0pt} \def\eqalign#1{\,\vcenter{\openup1\jot
\caja   \ialign{\strut \hfil$\displaystyle{##}$&$
\displaystyle{{}##}$\hfil\crcr#1\crcr}}\,} \newif\ifdtup

\def\frac#1#2{ {{#1} \over {#2} }}

\def\half{\mbox{\small $\frac{1}{2}$}}

\def\ie{\hbox{\rm i.e. }}

\def\abs#1{\left| \: #1 \: \right|}%
\def\bom#1{\mbox{\boldmath$#1$}}
\def\beq{\begin{equation}}
\def\eeq{\end{equation}}
\def\re#1{(\ref{#1})}
\def\ee{$e^+e^-\;$}
\def\as{\alpha_S}
\def\asb{\bar \alpha_S}

\def\bk{\bom {k} }
\def\bq{\bom {q} }
\def\bkq{\abs{\bom{k}+\bom{q}}}
\def\om{\omega}
\def\ga{\gamma}
\def\tga{\tilde \gamma}
\def\tchi{\tilde \chi}
\def\tom{\tilde \om}
\def\de{\delta}
\def\De{\Delta}
\def\cF{{\cal F}}
\def\cA{{\cal A}}
\def\Q_s{\mu}
\def\e{{\rm e}}
\def\max{{\rm max}}
\def\min{{\rm min}}

\def\np#1#2#3{Nucl.\ Phys.\ B#1 (19#3) #2}
\def\pl#1#2#3{Phys.\ Lett.\ #1B (19#3) #2}
\def\pr#1#2#3{Phys.\ Rev.\ D #1 (19#3) #2}
\def\prep#1#2#3{Phys.\ Rep.\ #1 (19#3) #2}

\def\zp#1#2#3{Zeit.\ Phys.\ C#1 (19#3) #2}
\begin{document}
\begin{titlepage}
\begin{flushright}
     hep-ph/9702418\\
     IFUM 552-FT\\
     February, 1997
\end{flushright}
\par \vskip 10mm
\begin{center}
{\Large \bf
Structure functions and angular ordering \\
at small $\bom{x}$}\footnote{Research supported
in part by the Italian MURST and the EC contract CHRX-CT93-0357
}
\end{center}
\par \vskip 1cm
\begin{center}
  \par \vskip 2mm \noindent
 {\bf G.\ Bottazzi, G.\ Marchesini, G.P.\ Salam, and M.\ Scorletti}\\
  \par \vskip 2mm \noindent
  Dipartimento di Fisica, Universit\`{a} di Milano,\\
  INFN, Sezione di Milano, Italy
\end{center}
\par \vskip 2mm
\begin{center} {\large \bf Abstract} \end{center}
\begin{quote}
 We compute the gluon distribution in deep inelastic scattering at
 small $x$ by solving numerically the angular ordering evolution
 equation.  The leading order contribution, obtained by neglecting
 angular ordering, satisfies the BFKL equation.  Our aim is the
 analysis of the subleading corrections.  Although not complete ---
 the exact next-to-leading contribution is not yet available --- these
 corrections are important since they come from the physical property
 of coherence of QCD radiation.  In particular we discuss the
 subleading correction to the BFKL characteristic function and the
 gluon distribution's dependence on the maximum available angle.
 Conformal invariance of the BFKL equation is lost, however this is
 not enough to bring the small-$x$ gluon distribution into the
 perturbative regime: although large momentum regions are enhanced by
 angular ordering, the small momentum regions are not fully
 suppressed.  As a consequence, the gluon anomalous dimension is
 finite and tends to the BFKL value $\gamma=1/2$ for $\alpha_S \to 0$.
 The main physical differences with respect to the BFKL case are that
 angular ordering leads to 1) a larger gluon anomalous dimension, 2)
 less singular behaviour for $x \to 0$ and 3) reduced diffusion in
 transverse momentum.
\end{quote}
\end{titlepage}

\section{Introduction}

Angular ordering is an important feature of perturbative QCD \cite{IR}
with a deep theoretical origin and many phenomenological consequences
\cite{AngOrd}.  It is the result of destructive interference: outside
angular ordered regions amplitudes involving soft gluons cancel.
This property is quite general.  It is present in both time-like
processes, such as \ee annihilation, and in space-like processes, such
as deep inelastic scattering (DIS).  Moreover it is valid in the
regions both of large and small $x$, in which $x$ is the registered
energy fraction in the \ee fragmentation function or the Bjorken
variable in the DIS structure function.  Due to the universality of
angular ordering one has a unified leading order description of all
hard processes involving coherent soft gluon emission.

Angular ordering is important in the calculation of multi-parton
distributions by resummation of powers of $\ln Q^2$, with $Q$ the hard
scale, and of powers of $\ln x$ or $\ln (1-x)$ for small or large $x$.
This is due to the fact that angular ordering defines the structure of
the collinear singularities and, to leading order, their relation to
the infra-red (IR) singularities for $x\to0$ or $x\to1$.  In
particular one finds that collinear singularities in the emitted
transverse momenta contribute both to $\ln Q$ and $\ln x$ or
$\ln(1-x)$.  This is because angular ordering implies ordering in the
emitted transverse momenta divided by the energies.

In this paper we start a systematic study of multi-parton emission in
DIS at small $x$.  The detailed analysis of angular ordering in
multi-parton emission at small $x$ and the related virtual corrections
has been done in Ref.~\cite{CCFM} (see also \cite{March}), where it was
shown that to leading order the initial-state gluon emission can be
formulated as a branching process in which both angular ordering and
virtual corrections are taken into account.  In this first paper we
study the fully inclusive gluon density which gives the structure
function at small $x$.  This gluon density is given by an inclusive
recurrence equation deduced from the small-$x$ coherent branching (the
CCFM equation).

In spite of the universality of angular ordering, the space-like and
time-like distributions in the small-$x$ region are profoundly
different even to leading order.  In \ee annihilation processes the
small-$x$ distributions are obtained by resumming the $\ln x$ powers
which come both from IR and from collinear singularities in the
angular ordered regions.  In DIS, angular ordering is essential for
describing the structure of the final state, but not for the gluon
density at small-$x$.  This is because in the resummation of singular
terms of the gluon density, there is a cancellation between the real
and virtual contributions. The only remaining collinear singularity is
the one originating from the first gluon emission.  As a result, to
leading order the small-$x$ gluon density is obtained by resumming
$\ln x$ powers coming only from IR singularities, and angular
ordering contributes only to subleading corrections.

The calculation of the gluon density by resummation of $\ln x$ powers
without angular ordering was done $20$ years ago \cite{BFKL} and
leads to the BFKL equation for $\cF(x,k)$, the gluon density at fixed
transverse momentum $k$, related to the small-$x$ part of the
gluon structure function $F(x,Q)$ by
\begin{equation}\label{cF}
F(x,Q)=\int dk^2 \;\cF(x,k)\Theta(Q-k)\,.
\end{equation}
In the moment representation the BFKL equation has solutions of the form
\begin{equation}\label{cFom}
x\cF(x,k)\;=\; \int\frac{d\om}{2\pi i}
\left(\frac{1}{x}\right)^{\om} \; \cF_{\om}(k) \,,
\;\;\;\;\;\;\;\;\;
\cF_{\om}(k) \;\sim\;
\frac{1}{k^2}\left(\frac{k^2}{k^2_0}\right)^\ga
\,,
\end{equation}
with $k_0$ an arbitrary constant and $\ga$ given by a solution of
the well known BFKL characteristic equation
\begin{equation}\label{chi}
1\;=\;\frac{\asb}{\om}\;\chi(\ga) \,,
\;\;\;\;\;\;\;\;\;
\chi(\ga)= 2 \psi(1)-\psi(\ga)-\psi(1-\ga)
\,,
\end{equation}
where $\asb = {C_A \as \over \pi}$ and 
$\psi$ is the logarithmic derivative of the gamma function.
The QCD coupling $\as$ is taken as a fixed parameter.
The renormalisation group dependence of $\as$ on a scale is an
effect which goes beyond this leading order contribution in
which one resums the powers $(\as/\om)^n$ for $\om \to 0$.
The next-to-leading order contribution, $\as (\as/\om)^n$,
is so far only partially computed \cite{NLO}.

In eq. \re{cF} $\ga$ plays a r\^ole analogous to that of the
gluon anomalous dimension, however its origin is not the
renormalisation group but conformal invariance,
which is a consequence of the absence of collinear singularities
and which implies that in $\cF(x,k)$ all regions of $k^2$
large or small are equally important.
This is reflected in the symmetry $\ga \to 1-\ga$ of the
characteristic function.

The gluon distribution in the angular ordered equation depends on an
additional variable $p$, which corresponds to the maximum available
angle for the initial state radiation. We denote by $\cA_\om(k,p)$ the
gluon distribution in this case. The solutions have a form similar to
\re{cFom}

\begin{equation}
\cA_\om(k,p) \sim \frac{1}{k^2}
		\left(\frac{k^2}{k_0^2}\right)^{\tga} G(p/k),
\end{equation}

\noindent where $G(p/k)$, describing the angular dependence, has a
structure typical of a form factor, vanishing for $p\to0$.  As in
\re{chi}, the gluon anomalous dimension $\tga$ is given by a modified
characteristic function $\tchi(\tga,\as)$ which depends also on
$\as$. Angular ordering breaks the conformal invariance of the gluon
density, so that the modified characteristic function is not symmetric
for $\tga\to 1-\tga$.

To leading order, i.e.\ for the leading powers $\as^n/\om^n$, the two
equations are equivalent, so that $\cA_\om(k,p)\to\cF_\om(k)$. We have
therefore that the angular dependence in $G(p/k)$ is a subleading
correction of order $\as$. The gluon anomalous dimension has the
expansion 

\begin{equation}\label{gaT}
\tga= \ga+\as \ga_1 +\cdots
\end{equation}

\noindent where the leading order result $\ga$, given by the BFKL
characteristic function, and the first correction $\ga_1$ are
functions of the ratio $\as/\om$. We have that $\as\ga_1$ is the part
of the next-to-leading correction of the gluon anomalous dimension
which comes from angular ordering.

The study of the differences between the two equations will be done by
analytical and especially by numerical calculations.  As we shall
discuss, the equation with angular ordering can be diagonalised only
partially thus numerical methods are needed.  The study of some of the
phenomenological features of the gluon density with angular ordering
has been done in Ref. \cite{Durham}.

In Sect.~2 we recall the main elements of the CCFM equation and its
relation to the BFKL one. We discuss how a hard scale enters.
We deduce some simple analytical properties and the behaviour of the
solution.
In Sect.~3 we discuss the numerical methods used.
In Sect.~4 we present the results.
Sect.~5 contains a summary of the main points and some
conclusions.

\section{Equation for the gluon density }

In this section we recall the basic elements for the small-$x$
coherent branching and the inclusive equation for the gluon density.

\begin{figure}
\begin{center}
\epsfig{file=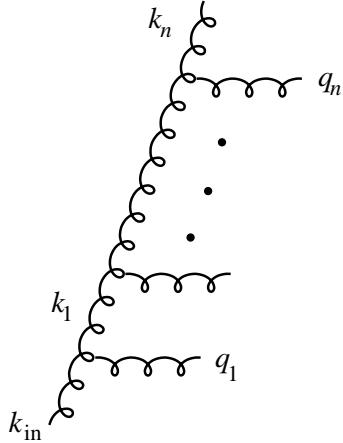}
\caption[]{Kinematic diagram for a DIS process at parton level:
 $k_{\rm in}$ is the incoming gluon, defined to have energy $E$; the
 $k_i$ are the exchanged gluons ($k_n$ is the gluon which undergoes
 the hard collision) and the $q_i$ are the gluons emitted in the
 initial state.}
\end{center}
\end{figure}

We start by considering the kinematic diagram for a DIS process at
parton level represented in fig.~1.  All partons involved are gluons
since gluons dominate the small-$x$ region.  The last exchanged parton
$n$ undergoes the hard collision at the scale $Q$.  For the exchanged
gluon $i$ we denote by $x_i$ and $\bk_i$ the energy fraction and
transverse momentum with respect to the incoming gluon.  Introducing
the energy ratio $z_i=x_i/x_{i-1}$, we have that $(1-z_i)x_{i-1}$ and
$\bom q_i$ are the energy fraction and transverse momentum of the
emitted gluon $i$. We shall use $k_i$ and $q_i$ also to denote the
moduli of the transverse momenta ($k_i = |\bk_i|$ and
$q_i=|\bq_i|$). We consider the region $z_i \ll 1$, which gives the
leading IR singularity.

The emission process takes place in the angular ordered region given
by $\theta_{i}>\theta_{i-1}$ with $\theta_{i}$ the angle of the
emitted gluon $q_i$ with respect to the incoming gluon $k_{\rm in}$.
In terms of the emitted transverse momenta $q_i$ this region
is given by
\begin{equation}\label{ao}
\theta_{i}>\theta_{i-1}\,,
\;\;\;\;\;\;\Rightarrow\;\;\;\;\;\;
q_{i} > z_{i-1} q_{i-1}
\,.
\end{equation}
The branching process given in \cite{CCFM} is accurate to leading
IR order and, at the inclusive level, does not require any collinear
approximation \cite{CFMO}.
The distribution for the emission of gluon $i$ is given by
\begin{equation}\label{dP}
d{\cal P}_i
=\frac{d^2q_i}{\pi q_i^2} \; dz_i\frac{\asb}{z_i}
\;\De(z_i,q_i,k_i)\;\Theta(q_i-z_{i-1}q_{i-1})
\,,
\end{equation}
where angular ordering \re{ao} is included.
The function $\De$ is the form factor which resums
important virtual corrections for small $z_i$
\begin{equation}\label{De}
\ln \De(z_i,q_i,k_i)=
-\int_{z_i}^1 dz' \;\frac{\asb}{z'}
\int\frac{dq'^2}{q'^2}\;\Theta(k_i-q')\;\Theta(q'-z'q_i)
\,.
\end{equation}

\noindent This form factor has a simple probabilistic interpretation. It
corresponds to the probability for having no radiation of gluons with
energy fraction $x'=z'x_{i-1}$ in 
the region $x_i<x'<x_{i-1}$, and with a  transverse
momentum $q'$ smaller than the total emitted transverse momentum $k_i$
and with an angle $\theta'>\theta_i$. Angles and momenta are related
by $q_i\simeq x_{i-1}E\theta_{i}$ and $q'\simeq x'E\theta'$ so that angular 
ordering gives $q'>z'q_i$. 
The two boundaries in $q'$ are due to coherence in the exchanged gluon
($k>q'$) and in the emitted one ($\theta'>\theta_i$).

One has

\begin{equation}\label{Dezqk}
\Delta(z_i,q_i,k_i) = \exp\left(-\asb 
 	\ln \frac{1}{z_i}\ln\frac{k_i^2}{z_i q_i^2}\right),
	\qquad k_i>q_i,
\end{equation}

\noindent so that this form factor has a double logarithmic
form, suppressing the radiation both for $z_i\ll 1$ and for the
emitted transverse momentum $q_i\ll k_i/\sqrt{z_i}$. 

 From the form in \re{Dezqk} and from the probabilistic interpretation,
we have that the function $\Delta$ plays a r\^ole similar to that of the
Sudakov form factor. However there are important differences. The
Sudakov form factor resums virtual corrections with IR singularities
due to soft emitted gluons, i.e.\ powers of $\ln (1-z)$, regularising
the $z\to1$ singularity in the splitting function. This implies that
when weighted with the energy fraction $z_i$, the usual branching with
the Sudakov form factor \cite{IR} can be normalised to unity,
corresponding to the $\omega=1$ energy sum rule.

The form factor \re{De} resums virtual corrections with IR
singularities due to soft exchanged gluons, i.e.\ powers of $\ln x$.
The branching \re{dP} cannot be easily normalised to unity for
$\omega=1$. However this normalisation is not relevant for our study
of small $x$. Note that $\Delta(z,q,k)$ depends not only on two
transverse momentum scales, as does the Sudakov form factor, but also
on the energy fraction. This extra dependence in the form factor is
one of the important features of DIS coherence at small
$x$. $\Delta(z,q,k)$ corresponds in the BFKL equation to the gluon
Regge form factor, which depends only on $k$, $z$ and a collinear
cutoff (see later).

Since the intermediate real and virtual transverse momenta are
bounded by angular ordering, no collinear cutoff is needed, except on
the emission angle of the first gluon.
However for $z_{i-1}\to0$ one exposes the collinear singularity
of $q_{i}\to 0$. Thus by integrating over the real and virtual transverse
momenta one generates powers of $\ln z_{i-1}$.
After integrating over the energy fractions $z_{i-1}$ one finds that
the general perturbative term is of the form
\begin{equation}\label{pert}
\frac{\as^n}{x}\ln^{n+\ell-1}x
\;\;\;\;\;\;\Rightarrow\;\;\;\;\;
\frac{\as^n}{\om^{n+\ell}}\,,
\;\;\;\;\;\;\;\;\;\;\;
\ell < n\,.
\end{equation}
Here each energy fraction integration gives a power of $\ln x$ while
each transverse momentum integration gives either $\ln x$ or $\ln k$.
We have therefore that the contributions with $\ell>0$ are obtained from
collinear singularities. Thus if collinear singularities cancel then
the leading $\ln x$ contributions are obtained only from IR
singularities, \ie for $\ell=0$.

In order to deduce a recurrence relation for the inclusive distribution
in the last gluon with fixed $x=x_n$ and $k= k_n$ one has to
introduce a transverse momentum $p$ associated with the maximum
available angle 
\begin{equation}\label{max}
\theta_n <\bar \theta
\;\;\;\;\;\;\Rightarrow\;\;\;\;\;
z_nq_n < p \simeq xE\bar\theta
\,,
\end{equation}
with $xE$ the energy of the last gluon, $k_n$, which undergoes the
hard collision at the scale $Q$. Then one defines the distribution for
emitting $n$ initial state gluons 
\begin{equation}\label{An}
\cA^{(n)} (x,k,p) \;=\; \int \prod_{i=1}^{n} \; d {\cal P}_i \;
\Theta(p-z_nq_n) \; \de(k^2-k_n^2) \; \de(x-x_n)
\,.
\end{equation}
The fully inclusive gluon density
\begin{equation}\label{A}
\cA(x,k,p) \;=\; \sum_{n=0}^{\infty} \; \cA^{(n)} (x,k,p)
\,,
\end{equation}
satisfies the following recurrence relation
\begin{equation}\label{A1}
\cA(x,k,p) \;=\; \cA^{(0)}(x,k,p) \;+\;
\int\frac{d^2q}{\pi q^2}\; \frac{dz}{z}
\;\frac{\asb}{z}\De(z,q,k)\;
\Theta(p-zq)\;\cA(x/z,\bkq,q)
\,,
\end{equation}
where the inhomogeneous term $ \cA^{(0)}(x,k,p) $ is the
distribution for no gluon emission.
This equation can be partially diagonalised by introducing the
$\om$-representation \begin{equation}
\cA_{\om}(k,p) \;=\; \int_0^1 dx \; x^{\om}\cA(x,k,p)
\,.
\end{equation}
One finds
\begin{equation}\label{A2}
\cA_{\om}(k,p) \;=\; \cA_{\om}^{(0)}(k,p) \;+\;
\int\frac{d^2q}{\pi q^2}\; dz\;\frac{z^{\om}\;\asb}{z}\De(z,q,k)\;
\Theta(p-zq)\;\cA_{\om}(\bkq,q)
\,.
\end{equation}
This equation cannot be further diagonalised in transverse
momentum since the kernel depends both on the total momentum $k$ and
on $q$ and $p$. Numerical studies are then necessary.

For the fully inclusive gluon density $\cA(x,k,p)$ there is a
cancellation between the collinear singularities which appear in the
real and virtual contributions of the kernel. To see this we convert
the recurrence relation into an inclusive form. By using the identity
\begin{equation}
\int_0^1 dz \; \frac{z^{\om}\;\asb}{z}\;\De(z,q,k)\; \Theta(p-zq)
=\frac{\asb}{\om}
\left\{ 1 \;-\; \int_0^1dz \;z^{\om}\;\frac{\partial}{\partial z}
\;\De(z,q,k)\;\Theta(p-zq)
\right\}
\,,
\end{equation}
one finds
\begin{equation}\label{A3}
\cA_{\om}(k,p) \;=\; \tilde \cA_{\om}^{(0)}(k,p) \;+\;
\frac{\asb}{\om}
\int\frac{d^2q}{\pi q^2}\;
\left[
\cA_{\om}(\bkq,q)
\;-\;
\Theta(k-q)\;\cA_{\om}(k,q_1)
\right]
\;+\;
\de_{\om}(k,p)
\,,
\end{equation}
where the inhomogeneous term is
\begin{equation}
\label{A3b}
\tilde \cA_{\om}^{(0)}(k,p) =
\cA_{\om}^{(0)}(k,p)
+\frac{\asb}{\om}\int\frac{dq^2}{q^2}\Theta(k-q)\cA^{(0)}_{\om}(k,q_1)
\,,
\end{equation}
and $q_1= \hbox{min} (q,p)$. The correction $ \de_{\om}(k,p) $ is given by
\begin{equation}\label{de}
\de_{\om}(k,p) \;=\;
\frac{\asb}{\om}
\int_p\frac{d^2q}{\pi q^2}\;
\cA_{\om}(\bkq,q)
\left[ \left(\frac{p}{q}\right)^{\om}\De(\frac{p}{q},q,k)\;-\;1\right]
\,,
\end{equation}
where the integration range is $q>p$.  In equation \re{A3} the first
term in the integral comes from the real emission contribution while
the second is due to the virtual correction. One sees explicitly that,
as in the BFKL equation, the real and virtual $q\to0$ collinear
singularities in the kernel cancel. Also $\de_\om(k,p)$ has no
collinear singularities since $q>p\neq 0$.

For $p\to\infty$ the term $\de_{\om}(k,p)$ vanishes and
the gluon density $\cA_{\om}(k,p)$ becomes independent of $p$.
In fact neglecting $\de_{\om}(k,p)$ and the $p$ dependence in $\cA_{\om}(k,p)$
one finds that \re{A3}
becomes the BFKL equation for the gluon density $\cF_{\om}(k)$
\begin{equation}\label{bfkl}
\cF_{\om}(k) \;=\; \tilde \cF_{\om}^{(0)}(k) \;+\;
\frac{\asb}{\om} \int\frac{d^2q}{\pi q^2}\;
\left[
\cF_{\om}(\bkq)
\;-\;
\Theta(k-q)\;\cF_{\om}(k)
\right]
\,.
\end{equation}
Neglecting the $p$-dependence in $\cA_{\om}(k,p)$ corresponds
to neglecting angular ordering.
To see this we modify the branching distribution in \re{dP}
and the virtual corrections \re{De} by neglecting angular ordering
so that the transverse momenta have no lower bound. To avoid
singularities we have to set a collinear cutoff $\mu$, which,
at the inclusive level, becomes irrelevant.
The modified branching distribution is given by
\begin{equation}\label{dP0}
\eqalign{
&
d{\cal P}_i^{(0)}
=\frac{d^2q}{\pi q_i^2} \; dz_i\frac{\asb}{z_i}
\;\De^{(0)} (z_i,k_i)\;\Theta(q_i-\mu)
\,,
\cr&
\ln \De^{(0)}(z,k)=
-\int_z^1 dz' \;\frac{\asb}{z'}
\int\frac{dq'^2}{q'^2}\;\Theta(k-q')\;\Theta(q'-\mu)
\,,
}
\end{equation}
obtained from \re{dP} and \re{De} by the substitution
$\Theta(q_i-z_{i-1}q_{i-1}) \to \Theta(q_i-\mu)$,
and $\Theta(q'-z'q) \to \Theta(q'-\mu)$ respectively.
This modification has no effect to leading order since the collinear
singularities cancel. Proceeding as for $\cA$, the gluon density
satisfies the following recurrence relation:
\begin{equation}\label{bfkl1}
\cF(x,k) \;=\; \cF^{(0)}(x,k) \;+\;
\int\frac{d^2q}{\pi q^2}\; \frac{dz}{z}
\;\frac{\asb}{z}\De^{(0)}(z,k)\;\Theta(q-\mu)\;\cF(x/z,\bkq)
\,,
\end{equation}
where the inhomogeneous term is related to the one in \re{bfkl}
as in \re{A3b}.
 From this modified branching one obtains
the BFKL equation \re{bfkl} in which the momentum has the cutoff $\mu$
which can be neglected.

\subsection{General properties of gluon distributions}
\label{sssGPGD}
In the following we discuss some of the properties of the gluon
distribution $\cA_{\om}(k,p)$ and its comparison with the BFKL
distribution $\cF_{\om}(k)$.
As mentioned before, at large $p$ the leading order contribution
to  $\cA_{\om}(k,p)$ tends to $\cF_{\om}(k)$.
We are interested in analysing the subleading corrections contained
in $\cA_{\om}(k,p)$ which are due to angular ordering.

We start by recalling some well known properties of the BFKL distribution.
The solution of \re{bfkl} is given by
\begin{equation}\label{bfkl2}
\cF_{\om}(k)=\int_{\frac12 - i \infty}^{\frac12 + i \infty}
\frac{d\ga}{2\pi i}
\;\frac{1}{k^2}\left(\frac{k^2}{k^2_0}\right)^{\ga}
\frac{\om f_0(\om,\ga)}{\om-\asb\chi(\ga)}
\,,
\end{equation}
where $k_0$ and the function $f_0$ are fixed by the inhomogeneous
term and $\chi(\ga)$ is the BFKL characteristic function \re{chi}.

Taking the initial condition
\begin{equation}\label{BFKLinit}
\tilde \cF_{\om}^{(0)}(k) \;=\;
\;\frac{1}{k^2}\left(\frac{k^2}{k^2_0}\right)^{\ga_0}
\,,
\end{equation}
with a given $\ga_0$ one has that the solution $\cF_{\om}(k)$
has the same form, and therefore the small-$x$ behaviour of
$\cF(x,k)$ is given by
\begin{equation}
x \cF(x,k) \sim
\;\frac{x^{-\om_c}}{k^2}\left(\frac{k^2}{k^2_0}\right)^{\ga_0}
\,,
\;\;\;\;\;\;\;\;\;\;
\om_c= \asb\;\chi(\ga_0)
\,.
\end{equation}
For a general initial condition the asymptotic behaviour of
$\cF_{\om}(k)$ for $k \gg k_0$ and for $k \ll k_0$ is given
by the expression \re{cFom} where $\ga$ is the solution of the
characteristic equation \re{chi} in the region $0<\ga<\half$
and $\half<\ga<1$ respectively.
The behaviour of $\cF(x,k)$  at small $x$ is determined by
the leading singularity of $\ga(\as/\om)$ in the $\om$-plane
which is at $\ga_c=\ga(\as/{\om_c})=1/2$ giving
$\om_c=\asb\chi(\half)=4\asb\ln2$.

We come now to discuss the properties for $\cA(x,k)$ by taking
solutions of the form 
\begin{equation}\label{cAom}
\cA_{\om}(k,p) =
\;\frac{1}{k^2}\left(\frac{k^2}{k^2_0}\right)^{\tga}
\;G(p/ k)
\,,
\end{equation}
where $\tga$ has to be specified and the function $G(p/k)$
takes into account angular ordering.

The equation for $G(p/k)$ is obtained by taking the derivative of
\re{A2} with respect to $p$:
\begin{equation}\label{dG}
p\;\partial_p\;G(p/k)=
\asb
\int_p\;\frac{d^2q} {\pi q^2}
\left(\frac{p}{q}\right)^{\om}\;\De(p/q,q,k)
\;G\left(\frac{q}{\bkq}\right)
\;\left(\frac{\bkq^2}{k^2}\right)^{\tga-1}
\,.
\end{equation}
We consider the case of $0<\tga<1$ and as a boundary condition we take
$G(\infty)=1$. This function depends on $\as$, $\om$ and $\tga$. 

If one takes the initial condition as for the BFKL case,
(see \re{BFKLinit})

\begin{equation}
	\cA^{(0)}_\om (k,p) = \frac{1}{k^2}
	\left(\frac{k^2}{k_0^2}\right)^{\tga_0 - 1}\Theta(p-k),
\end{equation}

\noindent with a given $\tga_0$, then the solution of the angular
ordering equation has the form \re{cAom} with $\tga=\tga_0$. In this
case $\tga$ is a free parameter independent of $\as$ and $\om$. 

The expression \re{cAom} is a solution of the homogeneous equation, as
in the BFKL case, provided that $\tga$ is given by the generalised
characteristic function which is obtained from \re{cAom} and \re{A3}
in the limit $p \to \infty$.

\begin{equation}\label{A5}
1=\frac{\asb}{\om}\tchi(\tga,\as)\,,
\;\;\;\;\;
\tchi=\int\frac{d^2q}{\pi q^2}
\left\{
\left(\frac{\bkq^2}{k^2}\right)^{\tga-1}
\;  G\left(\frac{q}{\bkq}\right)
-\Theta(k-q)\; G(q/k)\right\}
\,.
\end{equation}
There may be various solutions to this equation and we will consider
the leading one, i.e.\ that with the largest $\om$. In this case
$\tga$ is not an independent variable, but is a function of $\as$ and
$\om$.

 From these equations one finds that the leading order contribution to
$\cA(x,k,p)$ is given by $\cF(x,k)$.  First from \re{dG} one has that
the $p$-dependence of $G(p/k)$ is a subleading correction proportional
to $\asb$ without $1/\om$ enhancement.  Moreover for $G(p/k)\to 1$,
one has $\tchi(\tga,\as)\to \chi(\tga)$, the BFKL characteristic
function in \re{chi}.  We now list some properties of the angular
ordering function.

\noindent
{\it Behaviour of $G(p/k)$ for $p\gg k$}.
In the region  $q>p\gg k$ we have  $\De(p/q,q,k)=1$ and
$(\bk+\bq)^2 \simeq q^2$. From  \re{dG} we have
\begin{equation}\label{dG1}
p\;\partial_p\;G(p/k) \;\simeq\;
\asb \;G(1)\;\int_p\;\frac{d^2q} {\pi q^2}
\left(\frac{p}{q}\right)^{\om}\;\left(\frac{q^2}{k^2}\right)^{\tga-1}
\,.
\end{equation}
Since $\tga<1$, the derivative vanishes at large $p$ and one finds
\begin{equation}\label{G1}
G(p/k) \;\simeq\; 1
\;-\; \frac{\asb\;G(1)}{(1-\tga)(2-2\tga+\om)}
\;\left(\frac{p^2}{k^2}\right)^{\tga-1}
\,.
\end{equation}

\noindent
{\it Behaviour of $G(p/k)$ for $p\ll k$}.
By using
\[
\ln\;\De(p/q,q,k) \;=\; -\asb
\;\left[\ln^2(k/p)\;\Theta(k-p)\;-\;\ln^2(k/q)\;\Theta(k-q)\right]
\,,
\]
and  $(\bk+\bq)^2 \simeq k^2$ for $q \ll k$, from \re{dG} one obtains
\begin{equation}\label{dG2}
p\;\partial_p\;G(p/k) \;\simeq\; \asb \;e^{-\asb\ln^2(k/p)}
\;\left(\frac{p}{k}\right)^{\om}
\left[\int_p^k\;\frac{d^2q} {\pi q^2}\;\left(\frac{k}{q}\right)^{\om}
\;e^{\asb\ln^2(k/q)} \;G(q/k) \;+\; C(\om,\as)\right]
\, ,
\end{equation}
where $C(\om,\as)$ is a constant which is independent of $p$. 
We find then
\begin{equation}\label{G2}
G(p/k) \;\simeq\;
\frac{\asb\,C(\om,\as)}{\om}\;\left(\frac{p}{k}\right)^{\om}
\;e^{-\asb\;\ln^2(k/p)}
\,.
\end{equation}
This behaviour is similar to that of a Sudakov form factor. The
distribution is suppressed when the maximum angle $\bar\theta$
available for the initial state emission vanishes, more precisely for
$p\simeq xE\bar\theta$ much smaller than the total emitted momentum $k$.

\noindent
{\it Behaviour for $\tga\to0$ and $\as$ fixed}.
In this limit the angular ordering function assumes the perturbative
form, $G(p/k) \to \Theta(p-k)$. This is as one would expect from
\re{G1} and \re{G2}, and can be proved directly from \re{dG}.

\noindent
{\it Correction to the characteristic function}.  
The correction $\de\chi(\tga,\as)$ is given by eq.~\re{A5} in which
one substitutes $G(p/k)$ with $\de G(p/k)= 1-G(p/k)$. From \re{G1} one
has that $\de\chi=\chi(\tga)-\tchi(\tga,\as)$ is regular for
$\tga\to0$. From $\de\chi$ one obtains the subleading corrections
to the gluon anomalous dimension due to angular ordering.
By taking into account that $\ga$ is the solution of
$1=(\asb/\om) \chi(\ga)$ we can write the characteristic
equation \re{A5} in the form
\begin{equation}
\tchi(\tga,\as)=\chi(\ga)
\,.
\end{equation}
Expanding $\tchi(\ga,\as)$ around $\as=0$ we find
\begin{equation}\label{nlotga}
\tga= \ga + \as\ga_1 + \cdots\;,
\;\;\;\;\;\;\;
\ga_1= \left.
 -\frac{\partial \tchi(\ga,\as)}{\chi'(\ga)\partial \as}
 \right|_{\as=0}
\,.
\end{equation}

 \noindent From these equations we have that the first correction to
 $\tga$ is $\as\ga_1\sim \as^3/\om^2$. To prove this, observe that
 $\chi'(\ga)\sim \ga^{-2}$ for small $\ga$. Moreover, for $\as\to0$
 and $\ga$ fixed, we have $\de\chi(\ga,\as)\sim c\as$ where $c$ tends
 to a constant as $\ga\to0$.

 As we shall see from the numerical analysis, the characteristic
 function $\tchi(\tga,\as)$ decreases with $\tga$, reaches a minimum
 at $\tga_c<1$ (for reasonable $\as$), and then rises again. As in the
 BFKL case, we shall denote by $\tom_c$ the leading singularity in
 $\om$ which corresponds to the minimum of the characteristic function
 at $\tga=\tga_c$.

\section{Numerical methods}

\subsection{Evolution in rapidity}


 The angular ordered equation is solved by binning the function
 $\cA(x,k,p)$ in all three variables. To allow the coverage of a wide
 range of transverse and longitudinal momenta it is convenient to
 store the function on a grid of logarithmic variables $y=\ln(1/x)$,
 $\ln k$ and $\ln p$, where transverse momenta are in units of
 $k_0$. This allows us to go to very small $x$, and to cover the wide
 range of transverse momenta needed to correctly take into account the
 diffusion in $\ln k$. We solve for $\cA$ using the integral equation
 (\ref{A1}). From now on we will refer to the gluon density through
 the following function:

\begin{equation}
 A(y,k,p) = \e^{-y} \cA(\e^{-y},\bk,p).
\end{equation}

 \noindent The equation satisfied by $A$ is

\begin{equation}\label{Ay1}
 A(y,k,p) = A^{(0)}(y,k,p) +
	\asb \int_0^y dy' \int_{q_\min}^{q_\max}
	\frac{d^2q}{\pi q^2} \Delta(\e^{-y'},q,k)
	\Theta(\ln p + y' - \ln q) A(y-y', |\bk + \bq|, q),
\end{equation}

 \noindent where the limits on the $q$ integration, $q,\bkq>q_\min$
 and $q,\bkq< q_\max$, are introduced because of the finite extent of
 the grid in $\ln p$ and $\ln k$. Our approach is based on the fact
 that $A(0,k, p)=A^{(0)}(0,k,p)$. Then we attempt to determine $A$ on
 a grid in $y$, at points $y = i \delta y$, where $i$ runs from $1$ to
 some $i_\max = Y/\delta y$, $Y$ being the highest rapidity in which
 we are interested. Assuming that we know $A$ for all points on the
 grid up to $i\delta y$, then the procedure for evaluating the
 $(i+1)^{\rm th}$ point is the following: as a first approximation we
 set $A((i+1)\delta y,k,p) = A(i\delta y,k,p)$. This is put into the
 integral equation (\ref{Ay1}) to allow us to calculate a second
 approximation to $A((i+1)\delta y,k,p)$, which can itself be fed in
 to yield a still better approximation. This procedure is repeated
 until we have a stable value for $A((i+1)\delta y,k,p)$. Generally
 convergence is reached after about three or four steps. One can also
 aid the convergence by making a better first approximation (e.g.\ by
 taking into account the first derivative of $\cA$ with respect to
 $y$). In certain regions the form factor $\Delta$ varies very rapidly
 with $y'$, requiring the use of specially adapted integration weights
 to obtain the correct answer. In cases where $A$ varies with $y$ in a
 known rapid manner, that information can also be used. The BFKL
 equation is solved in a similar manner.

 Later in this article, for the purpose of obtaining precise
 information about $\tchi$, it will be necessary to go to extremely
 large rapidities --- $y\sim 100$, or equivalently $x \sim 10^{-50}$!
 At this point the $y'$ integrations, because of their large extent,
 become very slow, and also require one to store information about $A$
 at rapidities all the way from $0$ to $y$. This quickly becomes
 prohibitive both in terms of computing time (which scales as $y^2$)
 and memory requirements (one should bear in mind that for each $y$
 point, we are storing a large 2-dimensional grid in $\ln k$ and $\ln
 p$, as discussed below). Fortunately the integrand is dominated by
 small values of $y'$, because $A(y-y')$ is a function which decreases
 exponentially with $y'$ as does the form factor. This allows us to
 truncate the $y'$ integration; a limit of $y' < 4/\as$ is generally
 found to be adequate.

 The other component of the problem is the $d^2q$ integration. Given
 that we have stored $A(y,k,p)$ on a grid in $\ln k$ and $\ln p$, the
 task is that of obtaining a discretised kernel $K(i_k, i_{|k+q|},
 i_q)$ such that

\begin{equation}
 \int_{q_\min}^{q_\max} \frac{d^2q}{\pi q^2} A(y, |\bk + \bq|, q)
 = \sum_{i_q=-i_{q,\min}}^{i_{q,\max}} \;\;
	\sum_{i_{|k+q|}} K(i_k, i_{|k+q|}, i_q)
	A(y,\e^{i_{|k+q|}\delta \ell },\e^{i_q \delta \ell})
\end{equation}

 \noindent where $k=\e^{i_k \delta \ell}$ and $\delta \ell$ is the
 logarithmic spacing between grid points in $\ln k$ and $\ln p$. The
 sum over $i_{|k+q|}$ is the equivalent of the angular integral. The
 difficulty that arises is that in the region where $|\bk + \bq|$
 involves a significant cancellation, a small change in either $k$ or
 $q$ has a large effect on $\ln |\bk + \bq|$ --- so moving slightly
 away from a grid point defined by $i_k,i_q$, the result of the
 angular integral changes radically. The solution is to think of
 $i_k$, and $i_q$ not as grid points, but as extended regions in $k$
 and $q$ (and analogously for $i_{|k+q|}$), and when calculating the
 discretised kernel one must perform an average over the appropriate
 region. This is found to drastically reduce the discretisation
 errors.

 The main limits on the method described here are due to memory
 requirements resulting from storing the gluon density on a three
 dimensional grid. Generally the grid resolution
 parameters\footnote{$\delta y$ and $\delta \ell$ are kept equal to
 simplify the treatment of the $\Theta$-function in (\ref{Ay1}).} are
 $\delta y = \delta \ell \sim 0.1$, together with
 $q_\min\simeq10^{-6}k_0$ and $q_\max\simeq10^6k_0$. With these
 parameters one can determine $A$ to an accuracy of better than $1\%$.

 As done in section~\ref{characfn}, one can impose a dependence of the
 form $k^{2(\tga -1)} G^{(0)}(y,p/k)$ on $A^{(0)}(y,k,p)$. The
 $k^{2(\tga-1)}$ dependence remains in the solution $A$, so that it is
 no longer necessary to store the $k$ dimension of $A$.  This allows
 one to go to smaller bin spacings, increasing the accuracy, which is
 necessary when attempting precision studies of the angular ordered
 characteristic function $\tchi(\tga, \as)$. To perform the
 calculation at small $\tga$, one must take into account that the
 integral over the region of small $|\bk+\bq|$ needs to go to
 extremely small $|\bk+\bq|\ll \e^{-1/(2\tga)}$ (the integral is of
 the form $\int dx x^{2\tga -1}$). In principle this would require a
 prohibitively large number of bins --- but the problem has been be
 resolved by calculating analytically the contribution from the region
 below the lowest stored bin.

\subsection{Iterative method}

We have solved the recurrence relations derived from (\ref{A2}),
which in $\om$ space read:
\begin{equation}
\label{e33}
\cA_{\om}^{(r+1)}(k,p)
= \int \frac{d^2 q}{\pi q^2} 
\Gamma_{\om} (k,p,q)
\cA_{\om}^{(r)}\left(\bkq,q\right) 
\,,
\end{equation}
where the kernel $\Gamma$ is given by:
\begin{equation}
\label{e34}
\begin{array}{l}
\Gamma_{\om} (k,p,q) = \int_0^1 d z z^{\om} {\asb\over z} \Delta(z,k,q)
\theta(p - z q) 
\\[4pt]
= {\asb\over\om} \Big\{\left[q_1\over q \right]^\om -
\left[q_2\over q \right]^\om
+ {\sqrt{\pi} \om \over 2 \sqrt{\asb} }
\,\hbox{erfc} \left[ {\om\over2\sqrt{\asb} } + \sqrt{\asb} 
            \log {k\over q_2} \right]
\exp\left[{\om^2\over4\asb} + \om\log{k\over p} 
     + \asb\log^2 {k\over q_3} \right] \Big\} \,,
\end{array}
\end{equation}
with $q_1=\min(p,q)$, $q_2=\min(k,p,q)$ and $q_3=\min(k,q)$. The
corresponding equation for the BFKL distributions is
\begin{equation}
\label{e36}
\cF_{\om}^{(r+1)}(k)
= \Gamma_{\om} (k) 
\int \frac{d^2q}{\pi q^2} \theta (q-\mu) 
\cF_{\om}^{(r)} \left(\bkq\right) 
\,,
\end{equation}
where
\begin{equation}
\label{e37}
\Gamma_{\om} (k) = \int_0^1 d z z^\om {\asb\over z} \Delta^{(0)} (z,k) =
{\asb\over\om} 
{1\over 1 + {\asb\over\om} \log {k^2\over\mu^2} \theta (k-\mu) }\,. 
\end{equation}

The iterative method is not very efficient (in that it requires more
computer memory and CPU time than other methods), but has the
advantage that it closely mimics the physical branching process, thus
giving a simple way to evaluate final state (exclusive) quantities,
which will be examined in a future publication.

Our method is based on the truncated expansion of the various
distributions on a suitable basis of Chebyshev polynomials
(see also \cite{Durham}).
The expansion of the structure function \re{e33} reads:
\begin{equation}
\label{n1}
\cA^{(r)}_{\om} (k,p) \approx
w(k,p) \sum_{n=1}^{N} \sum_{m=1}^{M} 
a^{(r)}_{n m} V^{(N)}_n [t(k)] V^{(M)}_{m}[u(p)]  \,,
\end{equation}
where $w$ is a weight function, chosen to ensure convergence, the
functions $t$ and $u$ map the variables $k$ and $p$ (which range from
$0$ to $\infty$) onto the interval $[-1,1]$. The efficiency of the
method relies heavily on a suitable choice of the function $t$ and
$u$.  We have used the usual logarithmic mapping:
\begin{equation}\label{tkup}
t(k) = \left(\ln \sqrt{{k_{\max}\over k_{\min}}}\right)^{-1}
\ln {k\over \sqrt{k_{\min} k_{\max} } } \;,
\ \ \ \ \ \ 
u(p) = \left(\ln \sqrt{{p_{\max}\over p_{\min}}}\right)^{-1}
\ln {p\over \sqrt{p_{\min} p_{\max} } } \;,
\end{equation}
where $k_{\min}$, $p_{\min}$ and $k_{\max}$, $p_{\max}$ are
respectively the lower and upper cutoffs in the $k$ and $p$
variables. 

The basis functions are defined by 
\begin{equation}
V^{(N)}_n (t) = {2\over N} 
\left[ \sum_{i=1}^{N-1} T_{i}(t_n)T_{i}(t) + \frac12 \right],
\end{equation}
where $T_i$ are Chebyshev polynomials,
and are an orthonormal complete set on the points $\{t_n\}$,
\begin{equation}
t_n = \cos \left( {2N-2n+1\over2N}\pi \right) \;,
\ \ \ n=1,\ldots,N
\end{equation}
with the notable property that $V^{(N)}_n(t_m) = \delta_{n,m}$.

With these definitions,
the expansion coefficients are just 
the (rescaled) function values on the $N\times M$ rectangular grid 
$\{(k_n,p_m)\}$:
\begin{equation}
a^{(r)}_{n m} = {\cA^{(r)}_{\om} (k_n,p_m) \over w(k_n,p_m)} 
\end{equation}
where $t(k_n) = t_n , \ n=1,\ldots,N$
and $u(p_m) = u_m, \ m=1,\ldots,M$.

Inserting expansion \re{n1}, the recursion \re{e33} becomes
\begin{equation}
\label{n2}
a^{(r+1)}_{n m} = \sum_{n'=1}^N \sum_{m'=1}^M
L_{n m , n' m'} a^{(r)}_{n' m'} ,
\end{equation}
where the matrix
\begin{equation}
\label{n3}
L_{n m , n' m'} =
\int {d^2q\over\pi q^2} 
{w(\abs{\bk_m+\bq},q)\over w(k_n,p_m)}
\Gamma_{\om} (k_n,p_m,q)
V^{(N)}_{n'}[t(\abs{\bk_n+\bq})] V^{(M)}_{m'}[u(q)]
\end{equation}
must be evaluated only once, and then used to iterate the equation
as many times as desired.

We have checked the stability and the convergence of this method by
varying the functions $w$, $t$ and $u$, and the parameters $N$ and
$M$.  The method is weakly sensitive to the choice of $w$, as long as
the starting condition $\cA^{(0)}(k,p)\over w(k,p)$ is smooth, and
numerical convergence is ensured for the integral appearing in
\re{n3}.  We have used $w(k,p) = \cA^{(0)}(k,p)$ for the $\cA_\om$
distribution and $w(k)=\sqrt{{\mu^2\over k^2+\mu^2}}$ for the
$\cF_\om$ case.  The choice \re{tkup} allows us to obtain good
stability and fast convergence with $N$ and $M$ of the order of
$30-40$.

The number of iterations needed to obtain a stable solution depends
drastically on the value of $\om$.  Far from the $\om$-plane
singularity ($\om\approx 1$), $20-30$ iterations are sufficient to
obtain an accurate solution over all the $k$-range (the iteration
converges faster for smaller values of $k$).  On the other hand, the
required number of iterations increases dramatically as $\om$
approaches the critical value $\om_c$: with $\om-\om_c\approx 0.01$ we
need about $5000$ iterations to obtain a reliable result.

\section{Numerical results}

In this section we report the results obtained by solving
\re{A1} and \re{A2} for the gluon distributions
$\cA(x,k,p)$ and $\cA_{\om}(k,p)$ respectively.
We compare these results with those obtained from the
BFKL equation.

\subsection{Behaviour at small $\bom{x}$}
First we study $\cA(x,k,p)$ for a simple initial condition
\begin{equation}\label{inc}
\cA^{(0)}(x,k,p)=\de(1-x)\;\frac{1}{k}\de(k-k_0)\Theta(p-q_\min)
\,,
\end{equation}
 where $k_0$ sets the momentum scale. As a collinear cutoff for the
 first emission we take $q_\min$. This condition is not quite
 physical but is suitable for studying the general properties of the
 solution.  We first show that $\cA(x,k,p)$ becomes independent of $p$
 for $p \gg k$. Then we show that its behaviour for $x\to0$ is less
 singular than that of the BFKL gluon distribution.

\begin{figure}
\begin{center}
\epsfig{file=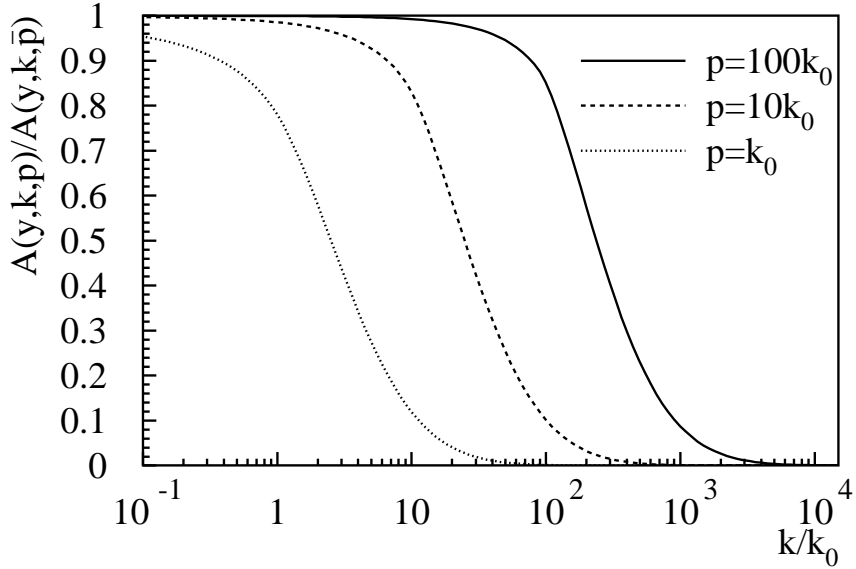, width=0.7\textwidth}
\caption[]{Plot of $\cA(x,k,p)/\cA(x,k,\bar p)$ against $k$ for
 different values of $p$; $x=\e^{-10}$, ${\bar p}=\e^{15}k_0$
 and $\as=0.2$.}
\end{center}
\end{figure}

\noindent
{\it $p$-dependence}.
In fig.~2 we plot $\cA(x,k,p)$ as a function of $k$ for
increasing values of $p$ and for fixed $x$ and $\as$.
As $p$ increases the gluon distribution becomes independent of $p$.
To show this we plot the ratio $\cA(x,k,p)/\cA(x,k,\bar p)$
with $\bar p$ in the asymptotic region.
As expected from the discussion in sect.~2 
the limiting value is
obtained first at low values of $k$.

\begin{figure}
\begin{center}
\epsfig{file=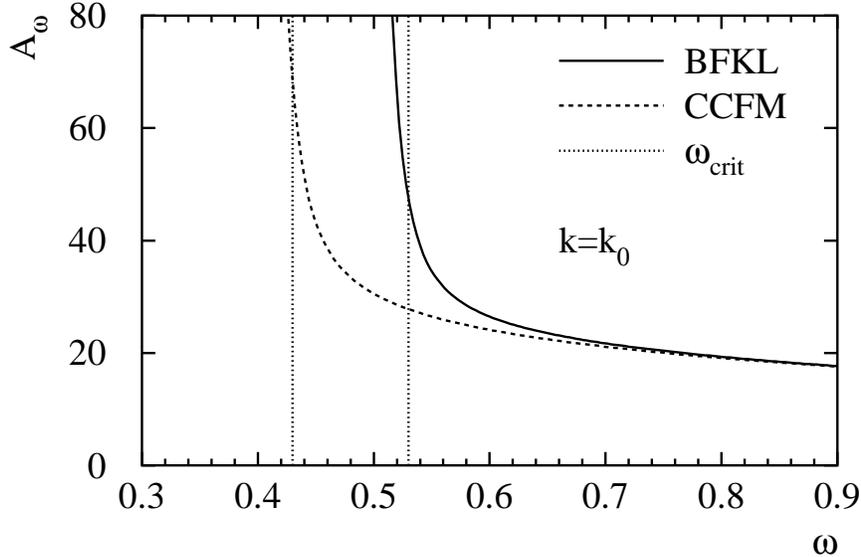,width=0.7\textwidth}
\caption[]{$\cA_\om(k,{\bar p})$ and $\cF_\om(k)$ for $\as=0.2$ and
 ${\bar p}=\e^{15}k_0$. For the BFKL case the dotted line corresponds
 to the exact singularity at $\om=\om_c=0.5295$. For the CCFM case the
 dotted line corresponds to a singularity at $\om=\tom_c=0.4301$, as
 determined in section~\ref{characfn}.}
\end{center}
\end{figure}

\noindent
{\it $\om$-plane singularity}.
 In fig.~3 we plot the two distributions $\cA_{\om}(k,\bar p)$ and
 $\cF_{\om}(k)$ as a function of $\om$ at fixed $k=k_0$ and $\as$.
 The value of $\bar p$ in $\cA_{\om}(k,\bar p)$ is in the asymptotic
 region.  From \re{bfkl2} one has that $\cF_{\om}(k)$ diverges at the
 singular point $\om_c=\asb \chi(1/2)=4\ln 2\asb$, the minimum of the
 BFKL characteristic function at $\ga=1/2$.  The gluon distribution
 $\cA_{\om}(k,\bar p)$ has a singularity at a value\footnote{The value
 quoted here is actually the one determined in section~\ref{characfn},
 which has a higher precision than that obtained by examining the
 position of the singularity of $A_\om$.} $\tom_c\simeq 0.4301$
 smaller than $\om_c\simeq0.5295$, the BFKL value.

 For the BFKL case the numerical distribution actually tends to
 diverge at a value of $\om$ which is $2-3\%$ smaller than $\om_c$.
 This is due to the presence of lower and upper limits, $q_\min$ and
 $q_\max$, in the $q$ and $k$ grid used for the numerical calculation.
 It is known \cite{MFR} that if the transverse momentum range is
 finite, the $\om$-singularity is shifted to a lower value of the
 order $\om_c\to \om_c(1-\pi^2/\ln^2q_\min/q_\max)$. Similar behaviour
 has also been noted \cite{GPS} in the context of the dipole approach
 to small-$x$ physics \cite{Dipole}. For the values of $q_\min$ and
 $q_\max$ considered here one finds $\pi^2/\ln^2q_\min/q_\max \simeq
 0.01 $.  The reason for this shift is that BFKL diffusion from the
 edges of the grid modifies the shape of the $k$-distribution which in
 turn leads to a reduction in the observed power. With angular
 ordering, the diffusion is reduced (this will be discussed also in
 section~\ref{characfn}) and therefore the edges of the grid have a
 smaller effect.


\begin{figure}
\begin{center}
\epsfig{file=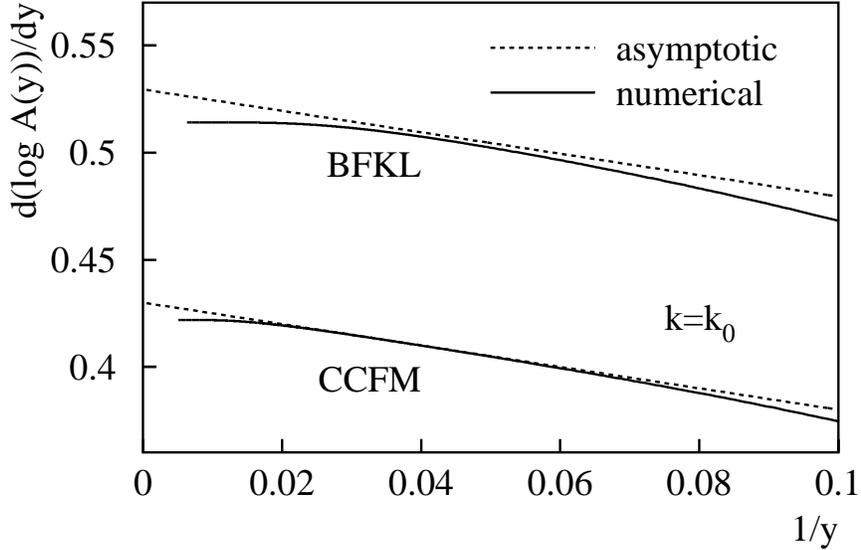,width=0.7\textwidth}
\caption[]{The effective power as a function of $1/y$; $\as=0.2$ and $p={\bar
 p}=3\times10^6$.}
\end{center}
\end{figure}

\noindent{\it Small-$x$ behaviour}.
 Another way to obtain the position of the $\om$-plane singularity
 consists in studying the small-$x$ behaviour of $\cA(x,k,\bar p)$ and
 $\cF(x,k)$. In particular in fig.~4 we plot in the small-$x$ region
\begin{equation}
  \frac{\partial}{\partial y} \ln x\cF(x,k) \simeq \om_c
-\frac{1}{2y}\,,
\;\;\;\;\;\;\;\;
y=\ln \frac 1 x
\,,
\end{equation}
 where $-1/2y$ is the first subasymptotic contribution.  One expects
 that there might be a similar kind of behaviour in the case of
 $\cA(x,k,\bar p)$, with $\om_c$ replaced by $\tom_c$.  This is indeed
 seen and as before we find $\tom_c<\om_c$.  The values for $\om_c$
 and $\tom_c$ agree with those found in the previous analysis in
 fig.~3.  In the BFKL case we have plotted the analytical result
 obtained for $q_\min \to0$ and $q_\max\to\infty$. With finite values
 of $q_\min$ and $q_\max$, increasing $y$, the width in $k$ of the
 solution increases until it is comparable to the extent of
 the finite grid, at which point the numerical curve flattens off. This
 is the same phenomenon that was noted previously for the shift of the
 singularity of $\cF_\om$. For the angular ordering case, we have
 plotted an analytical line analogous to the BFKL one, with the
 asymptotic power $\tom_c=0.43$ fitted to give agreement with the
 numerical results. The tailing off of the numerical curve at large
 $y$ is also observed in the angular ordering case, but it sets in
 later than for the BFKL curve --- consistent with the idea that
 diffusion is reduced by angular ordering so that the width of the
 solution approaches that of the grid only at larger $y$.


 We have also studied the $k$-dependence for large $y$.  In the BFKL
 case the numerical result fits well the expected behaviour $\sim
 1/kk_0$, corresponding to $\ga=\ga_c=1/2$.  With angular ordering we
 find $\cA(x,k,\bar p) \sim k^{2(\tga-1)}$ with
 $\tga=\tga_c\simeq0.61$. We shall analyse this behaviour in more
 detail later.

\subsection{Characteristic function}
\label{characfn}
 We now study the generalised characteristic function
 $\tchi$ and the corresponding angular ordering function
 $G(p/k)$.  Recall that for small $x$ the gluon distribution
 $\cA(x,k,p)$ has the asymptotic form
\begin{equation}\label{cAsym}
x\cA(x,k,p) \sim
\;\frac{x^{-\tom_c} }{k^2} \;\left(\frac{k^2}{k^2_0}\right)^{\tga-1}
\;G(p/k)
\,,
\;\;\;\;\;\;\;
\tom_c=\asb \tchi(\tga,\as)
\,.
\end{equation}
 To obtain $\tchi$ and the corresponding angular ordering function
 $G(p/k)$ as a function of $\tga$ and $\as$ we use the following
 method.  We solve equation \re{A1} by using a trial initial
 condition
\begin{equation}\label{ini}
\cA^{(0)}(x,k,p)= \frac{1}{k^2}\left(\frac{k^2}{k_0^2}\right)^{\tga}
\;\de(1-x)\;\Theta(p-k)
\,,
\end{equation}
 with a fixed value of $\tga$ and $\as$.  From the discussion in
 section~2 one has that asymptotically for $x\to0$, $\cA(x,k,p)$ has
 the form \re{cAsym} with the same $\tga$ as the inhomogeneous term
 \re{ini}. The $x$- and $p$-dependence of the initial condition is not
 important. Since the solution has the form of \re{cAsym} we only need
 to deal with $G$, which depends just on $p/k$. This means that one
 doesn't need to store the $k$ dependence of the solution and the
 finite extent of the grid no longer has a significant effect,
 drastically reducing the errors.

\begin{figure}
\begin{center}
\epsfig{file=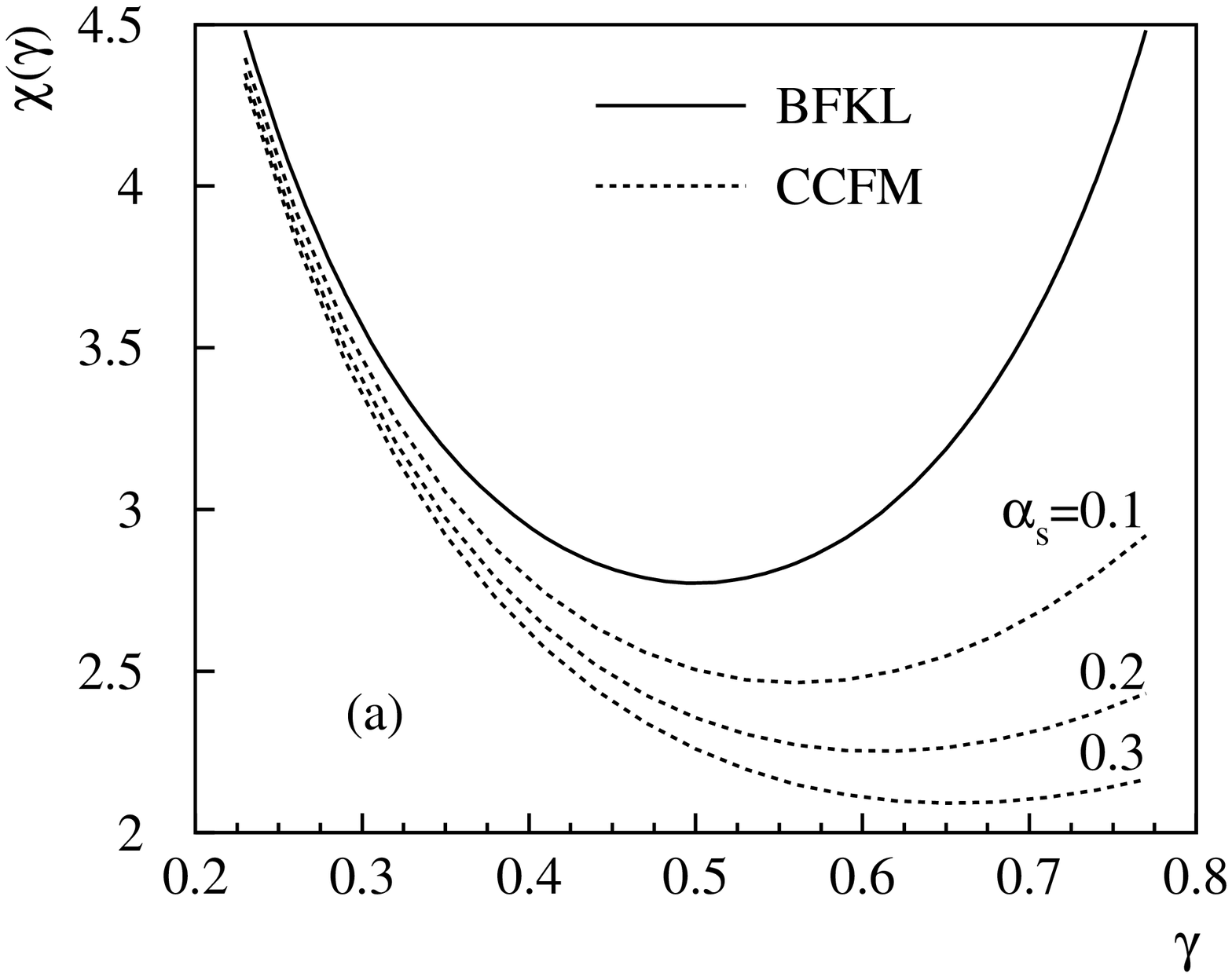,width=0.7\textwidth}\\
\epsfig{file=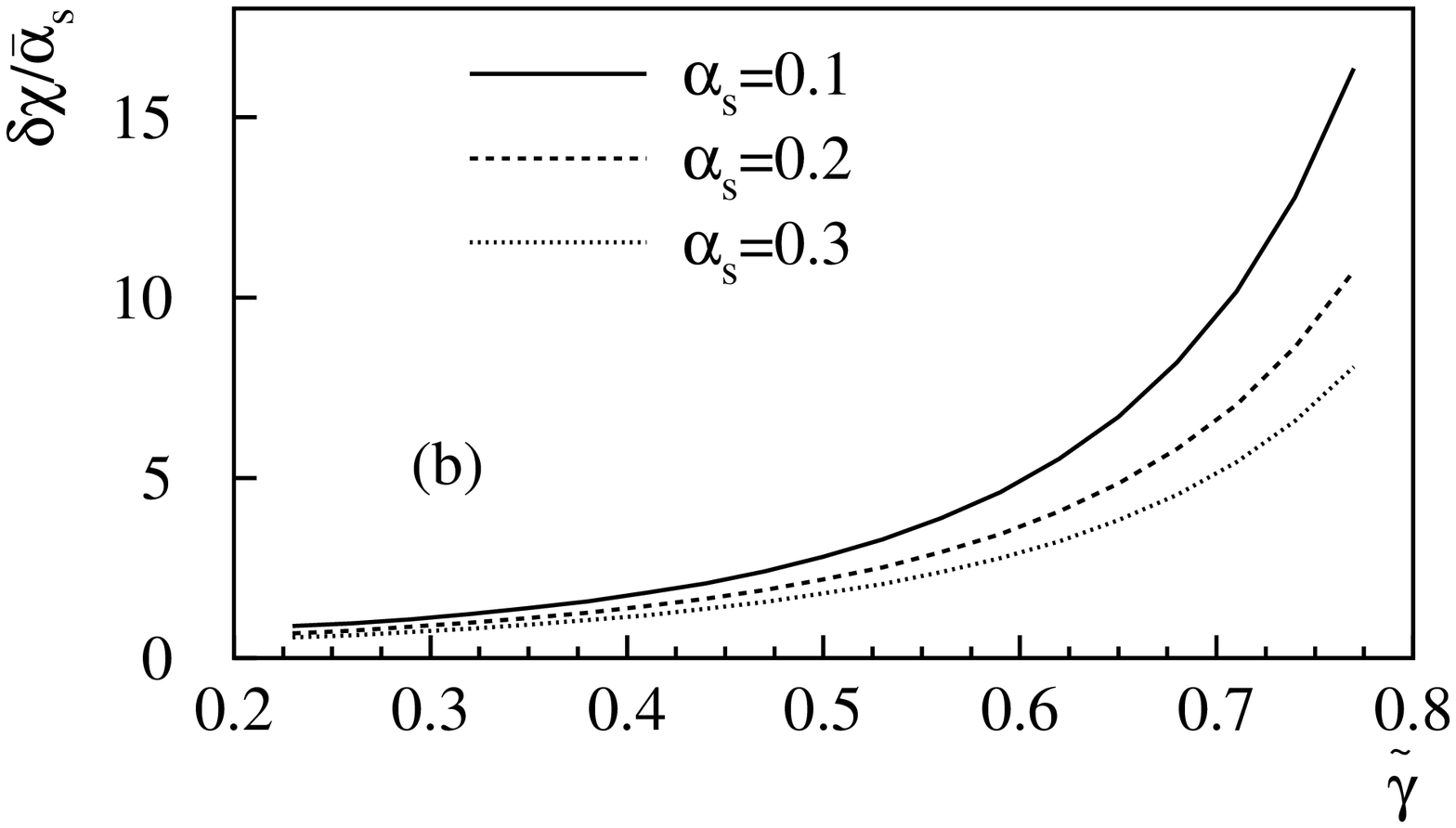,width=0.7\textwidth}
\caption[]{(a) The characteristic functions with and without angular
 ordering; $\tchi(\tga,\as)$ and $\chi(\ga)$ are plotted as functions
 of $\tga$ and $\ga$ respectively. (b) The difference,
 $\de\chi=\chi(\tga)-\tchi(\tga,\as)$, between the BFKL and angular
 ordered characteristic functions, divided by $\asb$.}
\end{center}
\end{figure}

\noindent
{\it Characteristic function $\tchi$}.
 To obtain $\tchi(\tga,\as)$ by using the initial condition \re{ini}
 we compute $\cA(x,k,\bar p)$ with $\bar p$ in the asymptotic region.
 By taking the small-$x$ limit we determine the intercept $\tom_c$
 with a high accuracy --- the relative error is of the order of
 $10^{-4}$ for much of the $\tga$ region. From $\tom_c$ we obtain the
 characteristic function
\begin{equation}
\tchi(\tga,\as)\;=\;\frac{\tom_c}{\asb}
\,,
\end{equation}
 as a function of $\tga$ and $\as$.  In fig.~5a we plot $\tchi$ as a
 function of $\tga$ for various $\as$.  We plot for comparison the
 BFKL characteristic function $\chi$.  We see that $\de \chi=
 \chi-\tchi$ is positive, increases with $\tga$, and increases with
 $\as$.  Moreover we find $\de \chi \sim \tga$ for $\tga\to0$ ($\asb$
 small and fixed) and $\de \chi \sim \asb$ for $\asb\to0$ ($\tga$
 small and fixed).  This agrees with our earlier observation in
 section~2 and implies that the next-to-leading correction to the
 gluon anomalous dimension coming from angular ordering is of order
 $\as^3/\om^2$. In fig.~5b we plot $\de\chi/\asb$ which shows that
 there are notable next-to-next-to leading corrections especially at
 large $\tga$.

\begin{figure}
\begin{center}
\epsfig{file=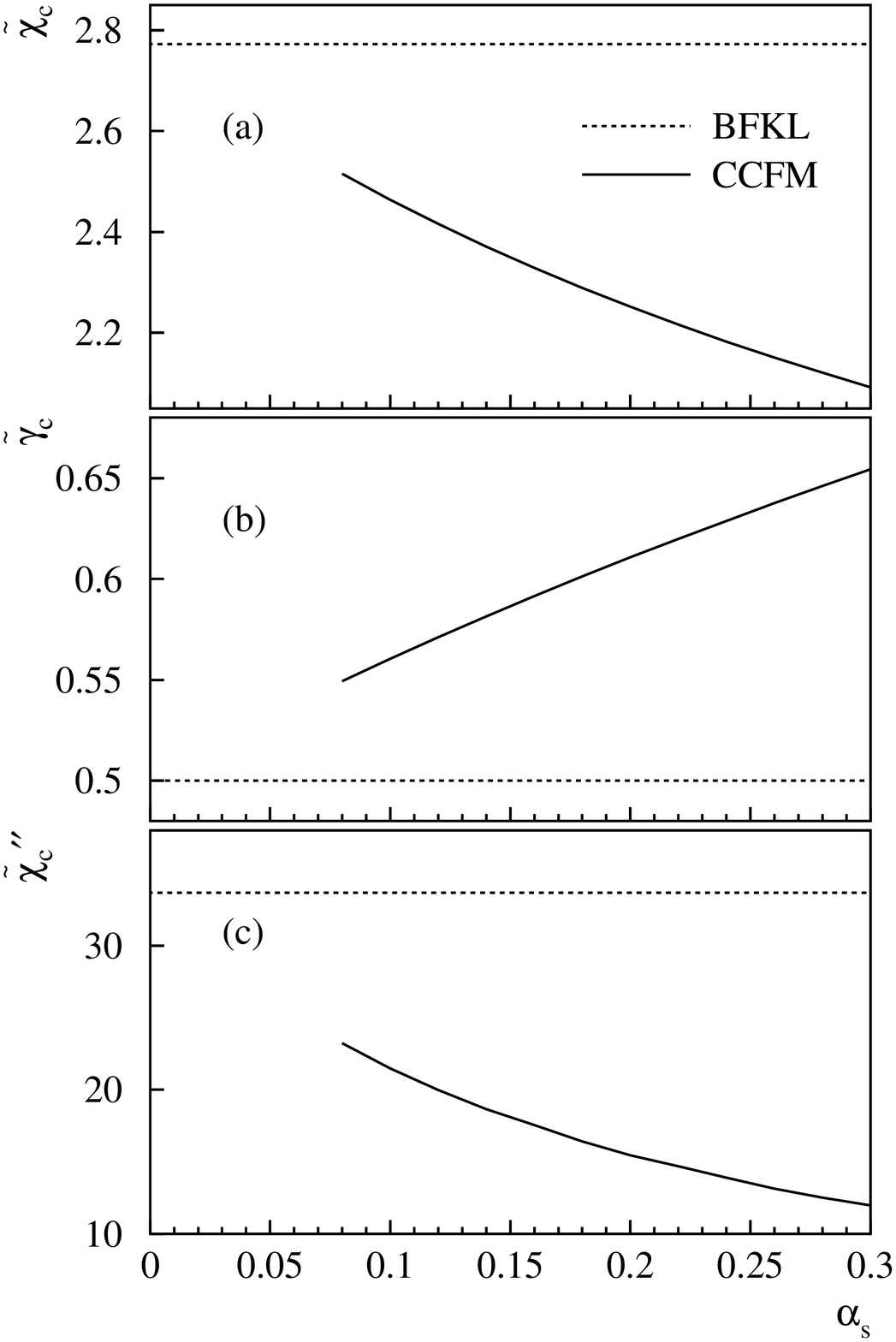,width=0.7\textwidth}\\
\caption[]{(a) The value of the minimum of the characteristic function,
 $\tchi_c$, as a function of $\as$. (b) The position of the minimum of
 the characteristic function, $\tga_c$, as a function of $\as$.
 (c) The second derivative of the characteristic function,
 ${\tchi_c}''$, at its minimum, as a function of $\as$.}
\end{center}
\end{figure}

 The symmetry for $\ga \to 1-\ga$ of the BFKL characteristic function
 is not valid for $\tchi$.  Recall that this symmetry is based on the
 fact that the regions of small and large $k$ are equally important.
 In the CCFM case, angular ordering favours instead the region of
 larger $k$. However, small values are still accessible. Therefore the
 function $\tchi$ decreases faster than $\chi$ for
 increasing $\tga$, but, after a minimum at a point $\tga_c$ larger
 than the BFKL value $1/2$, $\tchi$ increases again.  In fig.~6a and
 6b we plot as a function of $\as$ the values $\tchi_c$ and $\tga_c$
 with $\tchi_c$ the minimum of $\tchi$ and $\tga_c$ its position.  As
 expected the differences compared to the BFKL values $\chi_c=4\ln 2$
 and $\ga_c=1/2$ are of order $\asb$.  These results are consistent
 with the asymptotic solution in fig.~4. From our determination here,
 we obtain a very accurate estimate of the position of the singularity
 in $\om$: for $\as=0.2$, we find $\tom_c\simeq0.4301$ (the
 corresponding BFKL value is $\om_c\simeq0.5295$) and
 $\tga_c\simeq0.6106$.

 Figure~6c shows the second derivative, ${\tchi_c}''$, of the
 characteristic function at its minimum; this quantity is important
 phenomenologically because the diffusion in $\ln k$ is inversely
 proportional to the square root of ${\tchi_c}''$. From the figure,
 one can see therefore that the inclusion of angular ordering
 significantly reduces the diffusion compared to the BFKL case.

\begin{figure}
\begin{center}
\epsfig{file=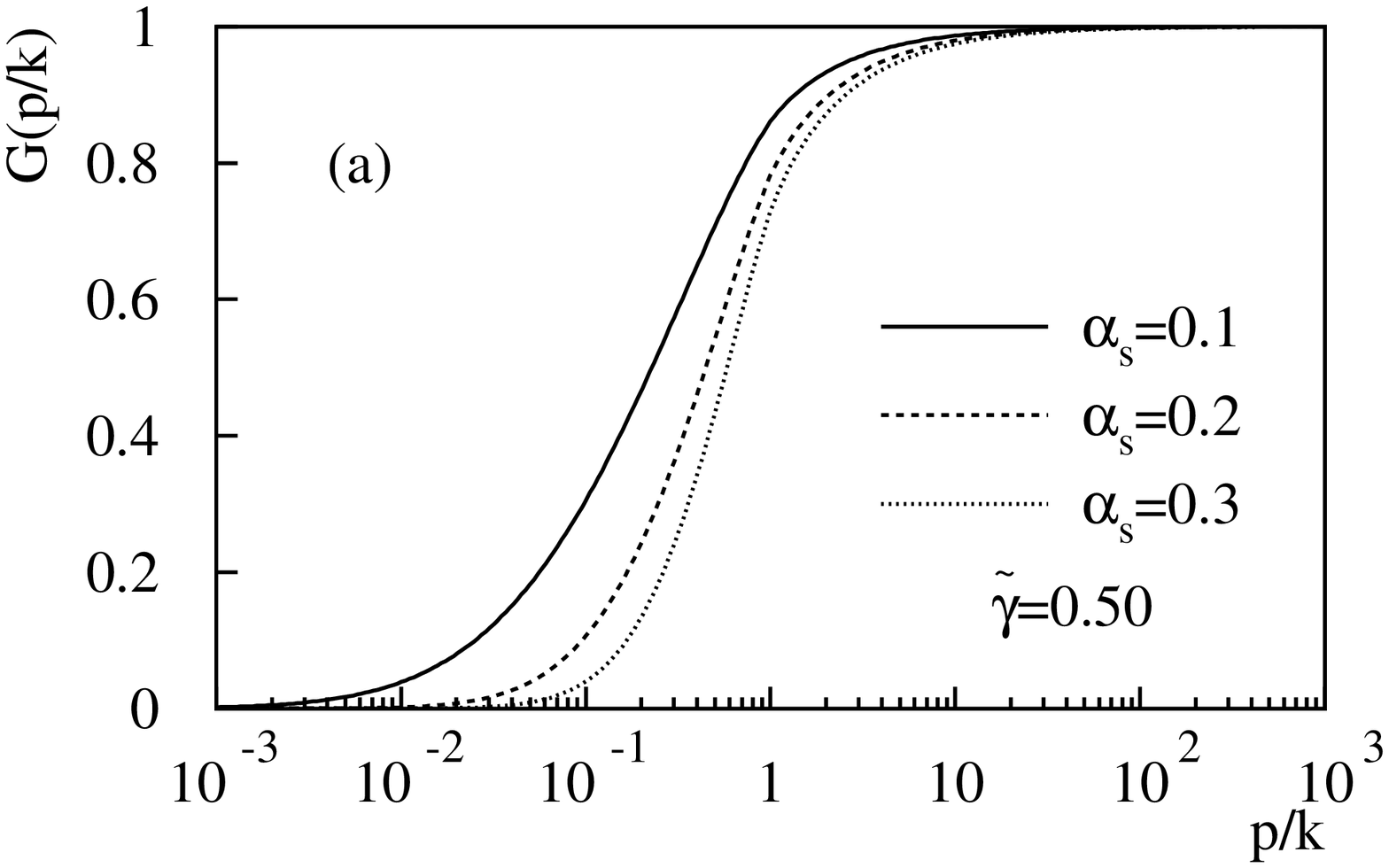,width=0.7\textwidth}\\
\epsfig{file=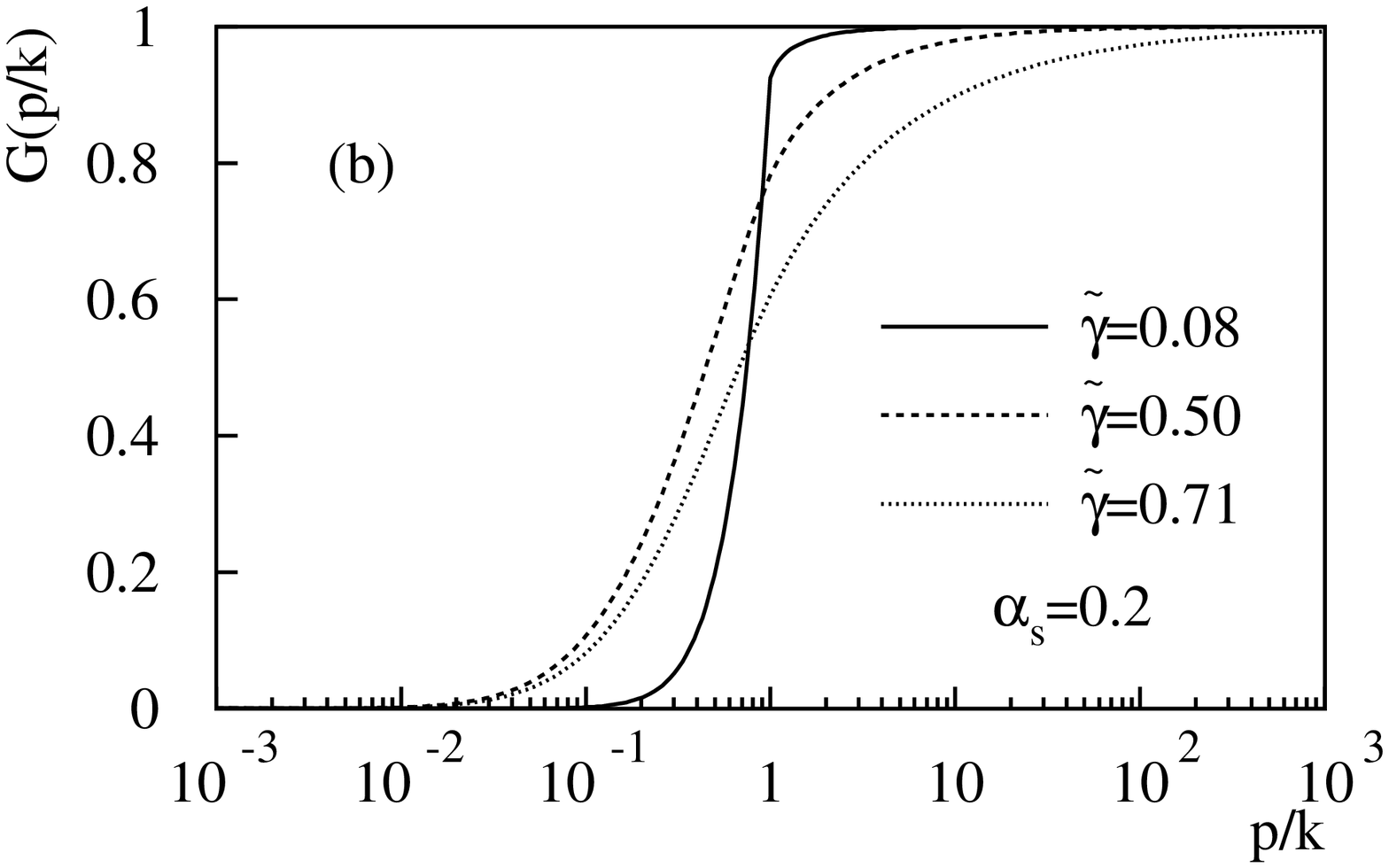,width=0.7\textwidth}\\
\caption[]{The angular ordering function $G(p/k)$: (a) for a range of
 $\as$, (b) for a range of $\tga$.} 
\end{center}
\end{figure}

\noindent
{\it Angular ordering function}.
 By using the trial initial condition \re{ini} and taking the
 small-$x$ limit for $\cA(x,k,p)$ we compute the angular ordering
 function $G(p/k)$ at the given value of $\tga$ and $\asb$ (see
 eq.~\re{cAsym}).  In fig.~7 we plot $G(p/k)$.  The behaviours of
 $G(p/k)$ for small and large $p$ (see \re{G1} and \re {G2}) are as
 expected. In particular, we note that as $\as\to 0$ (fig.~7a)
 $G(p/k)$ tends slowly to become $1$ everywhere; as $\tga\to 0$,
 $G(p/k)$ tends towards the function $\Theta(p-k)$.

\section{Conclusions}

 In this paper we have studied the contributions to the subleading
 corrections of the small-$x$ gluon density which are due to angular
 ordering.  Since they are based on the physical property of QCD
 coherence, one expects that they are among the important corrections.
 Another important subleading contribution, that which fixes the scale
 of the running coupling \cite{NLO}, is not included in our study.  The
 calculation has been done mostly by numerical methods, which prove to
 be quite reliable.  In future papers they will be extended to the
 study of associated distributions \cite{LMRW,Bart,Exp} for which angular
 ordering is relevant already to leading level.

 Our main results are summarised in figs.~5-7 in which we plot the
 generalised characteristic function $\tchi(\tga,\as)$ and the angular
 ordering function $G(p/k)$.  From these plots we have studied the
 subleading corrections $\de
 \chi(\tga,\as)=\chi(\tga)-\tchi(\tga,\as)$ and $\de G(p/k)
 =G(\infty)-G(p/k)$.  We find that $\tchi(\tga,\as)$ decreases with
 $\tga$ faster than the BFKL characteristic function, it has a minimum
 at $\tga_c$ which is larger than $\ga_c=1/2$, the BFKL critical
 point, and it rises again at larger $\tga$.

 The angular ordering function $G(p/k)$ has the structure of a typical
 form factor: it vanishes when the maximum available angle
 $\bar\theta$ vanishes, i.e.\ for $p\simeq xE\bar\theta$ much smaller
 than $k$, the total emitted momentum.

 The BFKL symmetry $\ga\to 1-\ga$ is lost since conformal invariance
 is broken by angular ordering.  The physical basis of conformal
 invariance is that in the BFKL equation the regions of small and
 large momentum are equally important.  The coherent branching instead
 tends to evolve toward large momenta.  However, at each intermediate
 branching, the region of vanishing momentum is still reachable for
 $x\to0$.  Within the angular ordering formulation, this effect has
 been discussed also in \cite{LMRW}.

 The fact that during the branching the intermediate momentum could
 vanish implies that the evolution contains non-perturbative
 components in an intrinsic way, not only in an initial boundary
 condition. It should be noted that in this non-perturbative region
 the distribution is non-singular (collinear singularities cancel), so
 that for any small but finite $x$ non-perturbative effects of the
 small-$k$ region are not too important. However they become very
 important for the asymptotic limit, $x\to0$.  As in the BFKL case the
 small-$k$ region generates a singularity in $\om$ at $\tom_c>0$ and
 an anomalous dimension $\tga_c$ which is non-vanishing with $\as$
 (see fig.~6b).

 By expanding $\tchi(\tga,\as)$ in powers of $\as$ at fixed $\as/\om$
 one obtains the part due to angular ordering of the next-to-leading
 correction $\as\ga_1(\as/\om)$ (see \re{nlotga}).  We have not
 obtained its analytical form but we have shown that the angular
 ordering correction in the small-$x$ limit starts with a power not
 smaller than $\as^3/\om^2$.

\vskip .3 true cm
\noindent{\bf Acknowledgements}\vskip .1 true cm
We are most grateful for valuable discussions with
M.\ Ciafaloni,
Yu.L.\ Dokshitzer,
A.H. Mueller
and
B.R.\ Webber.

\par 

\end{document}